# Ozone-based sequential infiltration synthesis of $Al_2O_3$ nanostructures in symmetric block copolymer

Jacopo Frascaroli*, Elena Cianci, Sabina Spiga, Gabriele Seguini, and Michele Perego*

Laboratorio MDM, IMM-CNR, Via C. Olivetti 2, 20864 Agrate Brianza, Italy.

**ABSTRACT:** Sequential infiltration synthesis (SIS) provides an original strategy to grow inorganic materials by infiltrating gaseous precursors in polymeric films. Combined with micro-phase separated nanostructures resulting from block copolymer (BCP) self-assembly, SIS selectively binds the precursors to only one domain mimicking the morphology of the original BCP template. This methodology represents a smart solution for the fabrication of inorganic nanostructures starting from self-assembled BCP thin films, in view of advanced lithographic application and of functional nanostructure synthesis. The SIS process using Trimethylaluminum (TMA) and $H_2O$ precursors in self-assembled PS-b-PMMA BCP thin films established as a model system, where the PMMA phase is selectively infiltrated. However the temperature range allowed by polymeric material restricts the available precursors to highly reactive reagents, such as TMA. In order to extend the SIS methodology and access a wide library of materials, a crucial step is the implementation of processes using reactive reagents that are fully compatible with the initial polymeric template. This work reports a comprehensive morphological (SEM, SE, AFM) and physico-chemical (XPS) investigation of alumina nanostructures synthesized by means of a SIS process using $O_3$ as oxygen precursor in self-assembled PS-b-PMMA thin films with lamellar morphology. The comparison with the $H_2O$-based SIS process validates the possibility to use $O_3$ as oxygen precursor expanding the possible range of precursors for the fabrication of inorganic nanostructures.

## I. Introduction

Sequential infiltration synthesis (SIS) technique allows the growth of inorganic materials into polymeric films by infiltrating precursors into the template from the gas phase. The selective binding of these precursors to only one domain of self-assembled block copolymer (BCP) domains provides the opportunity to produce inorganic functional nanostructures or hard masks for lithography applications.[1,2]

In BCP, the repulsive interaction between the constituent blocks leads to microphase separation into periodic domains with different morphologies and typical dimension below 50 nm.[3] The available morphologies for di-BCPs with two distinct polymeric units ranges from spheres to cylinders, to gyroids, to lamellae, depending on the volume ratio of the two blocks.[4,5] Infiltration of BCP thin films can readily provide the formation of aligned stripes with two-fold symmetry or ordered dots with hexagonal (six-fold) symmetry of inorganic materials.[3,6]

SIS is based on the stepwise controlled reaction of alternate precursors, originally developed for the atomic layer deposition (ALD) method, inside the polymeric matrix. Due to the chemical selectivity of the polymer blocks, the first precursor can be selectively incorporated in one of the polymer domains, while the inert chemistry of the other domain prevents trapping of the precursor. Subsequently, the injection in the reaction chamber of a suitable co-reactant completes the reaction of the inorganic material. This basic cycle can be iterated to saturate the original volume with inorganic material. After infiltration, the polymer matrix can be removed using a simple oxygen plasma or calcination step, leaving only the inorganic nanostructures.

The applicability of SIS in BCP thin films has been demonstrated for different inorganic materials, mainly oxides ($Al_2O_3$, $TiO_2$, $SiO_2$,[7] and $ZnO$[8]) but also W metal[7]. The main requirements are the low temperature, usually below 130°C to avoid polymer melting or mask degradation, and the ability of the precursors to easily diffuse inside the polymer matrix and selectively trap in only one of the polymer domains. Due to the high reactivity of Trimethylaluminum (TMA) at low temperature and its selective and controlled interaction with carbonyl groups present in the PMMA domains of self-assembled polystyrene-block-poly(methylmethacrylate) (PS-b-PMMA) BCP films,[9] the SIS process using TMA and $H_2O$ precursors in PS-b-PMMA BCP established as a model system for the investigation of the SIS technique. Moreover, one or a few SIS cycles of TMA and $H_2O$ have been used for seeding the growth of materials whose precursors would not normally have a sufficient interaction with carbonyl groups in PMMA domains.[7]

SIS has been mainly applied to enhance the etch resistance of organic BCP for pattern transfer of high aspect ratio features.[10–12] SIS of alumina in PS-b-PMMA templates has been applied for texturing large area Si surfaces, creating robust water repellent behavior.[13] Another notable example of possible implementation of this advanced nanofabrication technique is related to the synthesis of nanotextures for sub-wavelength photovoltaic applications.[14] However, this strategy also enables new fabrication approaches in which the functional nanostructures can be directly fabricated without the need of any pattern transfer process.[2] The formation of inorganic functional materials in periodic nanoarchitectures can trigger new applications in the field of nanophotonics,[15] in which the precise light-matter interaction provided by these structures can be exploited for light conversion and manipulation.[16,17] The selective infiltration in BCP scaffolds also provides unprecedented flexibility for the formation of tridimensional mesostructures with periodic arrangements[18], both inorganic and hybrid organic/inorganic, and hierarchical porous materials.[19,20]

In order to approach such potential applications, expanding SIS methodology to different materials, in particular metal oxides other than alumina, is required. However, some limitations on the SIS process are imposed by the low temperature processing constrain dictated by the use of polymers. Low temperature ALD processes are well known to often suffer of incomplete chemical reactions leading to inclusion of impurities in the growing film, of long reaction time, and even longer purge time to avoid chemical vapor deposition (CVD) reactions.[21] A way to enable low temperature ALD of metal oxides is the use of a highly reactive oxidizing precursor, that can efficiently react with the metal precursor.

Hitherto mainly $H_2O$ has been used as the oxygen precursor for the co-reactant step of the ALD reaction for oxide deposition, but at low temperature, low growth rates are reported due to insufficient energy to drive surface reactions, and furthermore $H_2O$ is difficult to purge away, leading to long process time. On the contrary, several reports highlight the benefits of $O_3$-based processes especially in low temperature ALD depositions, both in terms of material properties and of increased choice of precursors, thanks to the higher reactive property of $O_3$ against $H_2O$. For instance, $Al_2O_3$ films grown at low temperature using TMA and $O_3$ showed a reduced amount of hydrogen impurity and a shorter process time in combination with improved electrical properties compared to $Al_2O_3$ grown using $H_2O$.[22,23] As an example, in organic solar cells a thin encapsulation layer of $Al_2O_3$ deposited at low temperature by $O_3$-based ALD showed superior properties when compared to a layer deposited using $H_2O$.[24] Besides $Al_2O_3$, $O_3$ was applied for ALD deposition at relatively low temperature of other oxides, including $HfO_2$ and $VO_2$.[25–27] Therefore, the use of $O_3$ as co-reactant can be extended to SIS process, potentially enabling the infiltration of additional materials in BCP. This can pave the way for novel functional nanostructures and the creation of hard masks with etching selectivity to various substrates other than silicon.

In this work, we investigate the feasibility of an $O_3$-based process for the SIS of $Al_2O_3$ using TMA in nanostructured self-assembled PS-$b$-PMMA films. A perpendicular lamellar morphology is chosen to obtain aligned stripes of alumina with an aspect ratio higher than that obtained with parallel cylinders. As the physico-chemical analysis in combination with a morphology assessment of the produced nanofeatures is crucial in order to ascertain which applications are best suited for this fabrication approach, we investigate the morphology and physico-chemical properties of the $O_3$-based SIS $Al_2O_3$ and compare with a standard TMA and $H_2O$-based SIS process. Furthermore, $O_3$-based infiltration in continuous PMMA films is carried out to study the properties of the produced oxide in flat films in comparison with oxide nanostructures.

The results highlight the applicability of $O_3$ as oxygen precursor for the sequential infiltration of PMMA films and nanodomains with no noticeable polymer degradation up to ten SIS cycles. The morphology and chemical analysis showed overall comparable results between the two oxygen precursors, paving the way to the development of new $O_3$-based SIS processes with reactants requiring high chemical reactivity of the oxygen precursor.

## II. Experimental section
### II.A. Preparation of polymeric films.

The polymeric films were spin-coated on top of a (100) Si substrate with a native oxide layer of 1.7 nm. In order to clean the substrates and enhance the density of hydroxyl groups, the samples were treated in Piranha solution ($H_2SO_4/H_2O_2$, 3/1 vol. ratio) at 80°C for 40 min, dried under $N_2$ flow, and cleaned with 2-propanol in ultrasonic bath.

A PMMA homopolymer (Mn=21200 g mol$^{-1}$, PDI = 1.07) was purchased from Polymer Source Inc. and used as received. A solution of 8 mg in 1 mL toluene was spin-coated (30 s at 3000 rpm) on the cleaned Si substrate, obtaining a polymeric film with thickness of about 30 nm. The films were then thermally annealed at 290°C for 60 s in $N_2$ ambient by means of a Rapid Thermal Process (RTP) machine.

The perpendicularly oriented BCP morphology was obtained neutralizing the surface from any preferential interaction of the PS and PMMA blocks by a grafted film of random copolymer (RCP) prior to the BCP. A solution of 9 mg in 1 mL toluene of OH-terminated RCP (Mn = 13200 g mol$^{-1}$, PDI = 1.36, Styrene fraction = 0.58)[28] was spin-coated (30 s at 3000 rpm) on the cleaned Si substrates and thermally annealed in RTP at 310°C for 60 s in $N_2$ ambient to promote the polymer grafting to the surface. The samples were then washed in toluene to remove any non-grafted polymer chain. After washing in toluene to remove the ungrafted chains, the grafted RCP films had a thickness of ~7 nm.

A ~30 nm thick film of symmetric PS-$b$-PMMA BCP (Mn = 51 000 g mol$^{-1}$, PDI = 1.06, Polymer Source Inc.) was spin-coated (30 s at 3000 rpm) on the grafted RCP layer from a solution of 8 mg in 1 mL toluene. A perpendicularly oriented lamellar morphology was obtained upon phase separation promoted by a thermal treatment in RTP at 290°C for 60 s in $N_2$ ambient.[29,30]

### II.B. Sequential infiltration synthesis

The SIS process was performed in a Savannah 200 reactor (Cambridge NanoTech.) operating in a semi-static mode. In this particular operation mode, the outlet valve was closed during substrate exposure to precursors in order to maintain a constant environment and give time to the alternating precursors to diffuse inside the polymer film. TMA was used as Al precursor, while alternatively deionized $H_2O$ vapor or $O_3$ were adopted as oxidant. The $O_3$ was generated in a $O_2/N_2$ gas mixture with a concentration of 50 g N$^{-1}$ m$^{-3}$. The growth temperature was held fixed at 90°C during the whole infiltration process.

First, samples containing PMMA and self-assembled PS-$b$-PMMA BCP films were inserted in the ALD reactor and the chamber was evacuated down to a base pressure of 20 mTorr. The samples were then thermalized and excess moisture was removed keeping the samples under 100 sccm $N_2$ flow for 30 min at 90°C. A SIS cycle consisted first of TMA injection in the chamber, followed by 60 s exposure time and a 60 s purge step in 100 sccm $N_2$ flow. Afterward, the oxygen precursor was injected in the chamber followed by 60 s exposure time and a purge step of 300 s in 100 sccm $N_2$ flow. An oxygen plasma process (40 W, 0.7 mbar for 10 min) was performed to remove the organic component after SIS.

As a reference, an alumina film was deposited with 100 standard ALD cycles at 90°C on a Si substrate with native oxide using $O_3$ as oxygen precursor.[31]

After deposition, some samples were subjected to a thermal treatment performed at 400°C for 5 min in $N_2$ ambient in RTP.

### II.C. Film characterization

The film thickness and refractive index (n) were measured using a spectroscopic ellipsometer (SE, M-2000F, J. A. Woollam Co. Inc.) equipped with a Xe lamp, working in the spectral range between 250 nm and 1000 nm and with the beam angle of incidence with respect to the substrate normal of 75°. SE spectra of PMMA and self-assembled PS-b-PMMA BCP films, before and after SIS, as well as of Al2O3 films were fitted with a Cauchy dispersion relation and n was evaluated at 632.8 nm. SE data of alumina lamellae were modelled using the Bruggeman effective medium approximation (EMA).[32]

The structural morphology of the infiltrated BCP lamellae after the oxygen plasma step was characterized by field emission scanning electron microscopy (FE-SEM, SUPRA 40, Zeiss) using an in-lens detector and an acceleration voltage of 15 kV. The lamellar width was extracted from FE-SEM images averaging at least ten different lamellae, while the average lamellar spacing was determined from the first peak of the fast Fourier transformation of the FE-SEM images. The average lamellar height was determined from fit of SE data fixing the volume fraction occupied by the alumina lamellar features from FE-SEM images and the refractive index of $Al_2O_3$ extracted from SE of SIS $Al_2O_3$ in flat PMMA. A double layer model was used to take into account the $Al_2O_3$ layer formed in the infiltrated RCP layer at the interface between the self-assembled BCP film and the underlying substrate. The average height of the alumina lamellae produced using one single SIS cycle was also probed by scanning atomic force microscopy (AFM, Bruker Dimension Edge) using a topography probe (PPP-NCHR, Nanosensor) in tapping mode. This analysis gave results compatible within 1 nm with the estimation of the lamellae thickness obtained from the modelling of the SE data. For lamellar nanostructures produced using more than one SIS cycle the spacing between the nanostructures becomes comparable with the tip diameter and a reliable estimation of their morphology by AFM is prevented (Supplementary information S1).

The films were inspected by X-ray photoemission spectroscopy (XPS) on a PHI 5600 equipment using a monochromatic Al Kα source (1486.6 eV) and a concentric hemispherical analyzer. High-resolution spectra were acquired with a pass energy of 11.75 eV and a take-of-angle of 45°.

## III. Experimental results

### III.A. $Al_2O_3$ SIS in flat PMMA films

First of all, SIS using the $O_3$-based process was studied in flat PMMA films. In this way, it was possible to characterize the inherent oxide properties avoiding effects related to nanostructuring. Ten complete SIS cycles with alternating TMA and $O_3$ exposures were completed in ~30 nm thick PMMA films without any significant evidence of polymer degradation (*i.e.* increase of chamber pressure) during growth. After oxygen plasma, the residual thickness of the alumina film was 15.0 nm with n ~1.56 as extracted from SE data. Interestingly, after the oxygen plasma step, the thickness of the alumina layer infiltrated in a PMMA film with TMA and $H_2O$ was measured to be 14.5 nm with n ~1.54. The values of n are consistent with data estimated for $Al_2O_3$ films grown by standard ALD at T < 100°C using either $O_3$ or $H_2O$. The n values reported in the literature are comprised between 1.53 and 1.57.[23,33]

During the same growth with ten cycles using $O_3$, 1.6 nm of $Al_2O_3$ were grown on the pristine surface of a reference sample corresponding to a silicon substrate with native oxide. The formation of this $Al_2O_3$ film is attributed to the standard ALD reaction occurring at the $SiO_2$ surface during the repeated SIS cycles. Interestingly, for the same number of cycles, the final thickness of the $Al_2O_3$ film obtained by SIS in the PMMA template is significantly higher than the one obtained by a standard ALD process. This highlights the fundamental difference between the ALD process, which relies on surface functional groups, as opposed to SIS, which exploits the presence of reactive sites in the volume of the infiltrated polymer.

Figure 1 shows XPS high resolution spectra for the O 1s (left column), Al 2p (central column), and C 1s (right column) core level signals of the infiltrated PMMA films with ten SIS cycles. Data were acquired after the removal of the polymeric matrix by means of an oxygen plasma treatment. The spectra were normalized to the maximum value to simplify the comparison between the different samples. The binding energy scale was aligned with respect to the Al 2p core level position, according to data reported in the literature.[34] This calibration guarantees proper evaluation of the distance between the Al 2p and the O 1s core level signals, that is the most significant parameter to achieve information on the chemical composition of the $Al_2O_3$ films, since directly depends on the oxide stoichiometry [35] Actually the exact positions of the core level signals rigidly shift depending on the dielectric thickness due to charging effects during the XPS data acquisition[36]. In this regard, it is worth noting that rescaling the binding energy scale of the Al 2p and O 1s core level signals with respect to the C 1s signal corresponding to adventitious carbon (BE ~ 284.5 eV) leads to a similar result. However, the presence in the C 1s high resolution spectra of several contributions in addition to adventitious carbon is clearly highlighted in Figure 1 (right column). These additional C contributions derive from residual polymer in the alumina matrix and/or from the presence of surface carbonate species. Their intensities significantly vary depending on the specific sample, making the accurate determination of the effective position of the C 1s signal less accurate and consequently reducing the reliability of the energy scale calibration performed using this specific signal as a reference.

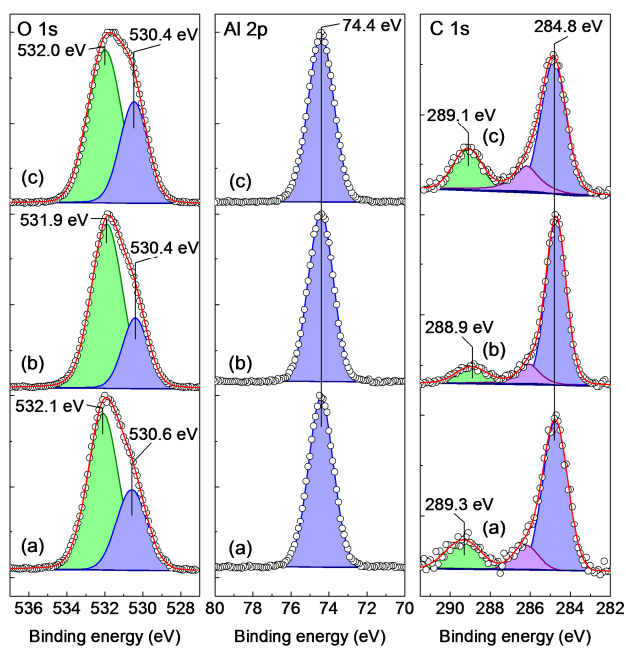

Figure 1. Normalized XPS O 1s, Al 2p, and C 1s high resolution spectra of the PMMA film after infiltration with ten SIS cycles using $O_3$ (a) or $H_2O$ (b) as oxygen precursors. (c) Normalized XPS spectra of an $Al_2O_3$ film deposited with 100 standard $O_3$-based ALD cycles.

As shown in Figures 1(a) and 1(b), both PMMA samples infiltrated using $H_2O$ and $O_3$ oxidants exhibits rather similar O 1s (left column) and Al 2p (central column) core level signals. In particular the Al 2p signals are characterized by a quite narrow FWHM of ~1.55 eV, which suggests the presence of one single component that corresponds to stoichiometric alumina.[34] Conversely, the O 1s core level signals are clearly asymmetric with pronounced shoulders at low BE. These considerations point towards the existence of at least two different contributions. Comparing the O 1s and Al 2p spectra of the infiltrated samples with the ones (Figure 1(c)) obtained from a reference $Al_2O_3$ sample grown by conventional ALD ($O_3$ precursor, 100 ALD cycles) at low temperature, the same general trend still holds. The only difference lies in a slightly higher intensity of the shoulder at low BE in the O 1s spectrum.

These high resolution spectra are equivalent to the ones reported in the literature for ALD films deposited using the $O_3$ precursor at 100°C[23] and $O_2$ plasma-enhanced ALD at 80°C[37] respectively. The broadening of the O 1s signal is typical of low temperature ALD processes and has been attributed to the presence of a O 1s component related to hydroxyl groups and carbonated species incorporated in the film during growth or adsorbed at the upper surface of the film[37,22]. The O 1s core level signal is reported to be the convolution of two Gaussian peaks with relative area of 70% and 30%, with the contribution at lower BE arising from Al-O bonds. By fitting the experimental data we determined the position of these two contributions, that are separated by ~1.5 eV. Considering the O 1s components at lower binding energy arising from Al-O bonds, the O 1s – Al 2p distance is determined to be 456.2 ± 0.1 eV in good agreement with data reported in the literature for stoichiometric $Al_2O_3$.[34]

Considering now the C 1s high resolution spectra (Figure 1, right column) we noticed the presence of several components. The main components of the C 1s signals are aligned at 284.8 ± 0.1 eV, which is perfectly compatible with the binding energy of the standard XPS signal that is conventionally attributed to the presence of adventitious hydrocarbon physisorbed on the sample surface. This asymmetric peak contains components also from C-O bonds at about 286 eV. In addition, a not negligible second component lies at a BE that is about 4 eV higher than the one of the main component This contribution to the C 1s spectrum is usually assigned to the presence of carbonate species.[38] The presence of carbonate species is attributed to carbon residues still remaining in the sample after the oxygen plasma treatment that is performed to remove the polymeric template. It is interesting to note that the highest carbonate signal is observed in the ALD sample as confirmed by time of flight secondary ion mass spectrometry (ToF-SIMS) depth profiles of the samples. This carbon contamination is explained by the incorporation of residual precursor that it is not completely desorbed during the growth process due the short purge step of only 20 s in the ALD process. Conversely the infiltrated PMMA samples were subjected to a much longer purge step of 60 s[39] that prevented incorporation of residual carbonates in the $Al_2O_3$ films.

According to reported data, SIS in PMMA polymer films using $O_3$ as oxygen precursor effectively leads to the growth of an alumina film with distinctive XPS features that are perfectly equivalent to those obtained with a standard ALD process performed at the same low temperature of 90°C. In addition, the refractive index is compatible with the one measured in alumina films grown by ALD at low temperature. Moreover, comparing alumina films produced by SIS in PMMA layers using $H_2O$ or $O_3$, from a chemical point of view the two precursors return similar results.

As previously highlighted the asymmetry of the O 1s core level signals suggests that a substantial amount of oxygen is not strictly related to the stoichiometric alumina but to hydroxyl groups or water physisorbed on the surface of the $Al_2O_3$ film due to air exposure. A post-deposition annealing can improve the oxide properties of alumina grown at low temperature.[40,41] Figure 2 reports the XPS high resolution spectra acquired after a thermal treatment at 400°C performed in $N_2$ atmosphere. Upon annealing, the high resolution spectra of the O 1s core level signals are radically changed. The asymmetry of the O 1s core level signal is reversed (Figure 2(a), left column), since the contribution at 1.5 eV above the BE of the $Al_2O_3$ shrinks to ~24% of the whole area in all the investigated samples. The component at low BE is now located at a distance of 456.4 ± 0.1 eV with respect of the Al 2p signal, which is consistent with data reported in the literature for stoichiometric $Al_2O_3$ grown at high temperature.[23,34,37] Moreover, the ratio between the areas of the low BE component of the O1 s signal and of the Al 2p signal confirms that this main peak is related to the $Al_2O_3$ stoichiometry.

After annealing, the chemical difference among the infiltrated flat PMMA films and the reference sample deposited by $O_3$-based ALD at low temperature is negligible. The refractive index of both PMMA infiltrated films significantly increases to ~1.69, which is close to the value reported for $Al_2O_3$ films grown by ALD at high temperature.[33,37] In addition, after thermal treatment the intensity of the overall C 1s core level signals as measured at the sample surface is almost halved, while the contribution associated with carbonate species has a similar area in the all inspected samples irrespective of the different growth processes. According to XPS and SE a thermal treatment at 400°C in inert ambient of alumina films produced by SIS yields a dielectric material that, from a chemical and optical point of view, is basically equivalent to the high quality $Al_2O_3$ films grown by ALD at high temperature.

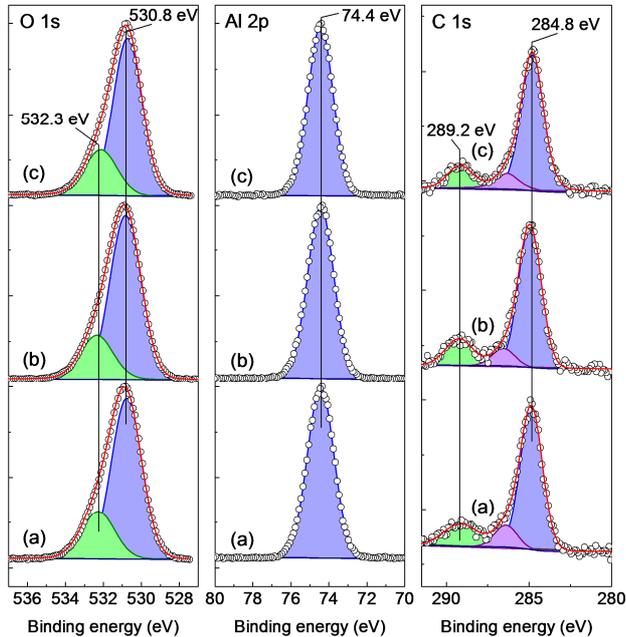

Figure 2. Normalized XPS O 1s, Al 2p, and C 1s high resolution spectra of the PMMA film after infiltration with ten SIS cycles using $O_3$ (a) or $H_2O$ (b) as oxidant reagent and post-deposition annealing at 400°C for 5 min in $N_2$ ambient. (c) Normalized XPS spectra of an $Al_2O_3$ film deposited with 100 standard $O_3$-based ALD cycles and the same post-deposition annealing.

### III.B. SIS in self-assembled PS-*b*-PMMA thin films

The perpendicularly oriented lamellar morphology for the infiltration of inorganic materials has not been thoroughly investigated as the parallel or perpendicularly oriented cylinder morphology. However, it provides the opportunity to obtain inorganic aligned stripes with aspect ratio higher than one, as opposed to parallel cylinders whose aspect ratio is intrinsically limited by the morphology of the BCP domains. A possible drawback of this choice is the presence of the grafted RCP film beneath the perpendicularly oriented lamellae. As previously explained this interfacial layer is required to neutralize the preferential wettability of the PS domain with respect to the PMMA domains on the $SiO_2$ surface, and thus to promote the perpendicular orientation of the lamellae with respect to the underlying substrate. However, the RCP film contains MMA monomers and consequently polar reactive sites for the TMA precursor. During SIS, this layer also infiltrates with the formation of an $Al_2O_3$ continuous layer at the interface between the BCP film and the substrate. Consequently, in case of pattern transfer, a breakthrough process with a specific dry etching chemistry is required to remove this layer and expose the Si substrate. A detailed knowledge of the infiltrated thickness as a function of the number of SIS cycles is required.[13,42]

For this reason, we performed a systematic investigation of the $O_3$-based SIS process in grafted RCP films with a thickness of ~7 nm and we afterward removed the residual polymer by plasma treatment. These grafted RCP films correspond to those deposited on the silicon substrate for surface neutralization before the BCP film deposition and self-assembly. After the first SIS cycle, the formed alumina layer measured 1.1 ± 0.1 nm, while after ten consecutive SIS cycles the alumina reached a thickness of 5.9 ± 0.1 nm. The plot of the infiltrated RCP thickness as a function of the number of SIS cycles is reported in Supplementary Information S2.

A viable route to mitigate the issue related to RCP layer hardening is to use ultrathin RCP grafted layers. By adjusting the molecular weight of the RCP, an effective surface neutralization for an RCP thickness as low as 2 nm was obtained.[43] Moreover, it was recently demonstrated that the RCP layer can be beneficial for the BCP self-assembly process as it acts as a reservoir for solvent during the annealing process.[44] The residual solvent trapped in RCP layers can enhance the lateral order and provide a way to overcome the topological constraints of the lamellar morphology.[45] From a general point of view it is interesting to note that in the case of cylinder forming PS-*b*-PMMA thin films parallel oriented with respect to the substrate no random copolymer is required to neutralized the substrate. Nevertheless, the formation of a ~1 nm thick $Al_2O_3$ layer at the interface between the BCP film and the underlying substrate is observed due to the reaction of TMA with the reactive sites that are present on the $SiO_2$ surface. Even in this case specific procedures are necessary to open the mask and remove this continuous $Al_2O_3$ layer prior to start the transferring of the pattern from the $Al_2O_3$ template to the underlying silicon substrate.[46]

The two SIS processes with the different $O_3$ and $H_2O$ oxygen precursors were performed in self-assembled PS-*b*-PMMA BCP thin films with spatially separated PS and PMMA lamellar domains. The morphological and chemical evolution of the $Al_2O_3$ nanostructures during the infiltration process was investigated as a function of the number of SIS cycles, since this can provide an additional degree of freedom for tuning the dimension of the inorganic features.

The adopted BCP template with perpendicularly oriented lamellae consists of aligned stripes of alternating PS and PMMA domains. The height of the lamellar domains is determined by the film thickness (30 nm), while the center-to-center distance (hereinafter called spacing for brevity) between adjacent lamellae is dictated by the molecular weight of the BCP and is settled at 26 nm for the specific BCP under investigation. The lamellar morphology of the phase separated BCP requires the adoption of a symmetric BCP with similar molecular weight for the two blocks. Consequently, the average size of the PS and PMMA domains corresponds to about 13 nm, i.e. half of the spacing. The thermal treatment for the BCP self-assembly was optimized in order to obtain a periodic ordered arrangement over reasonably large areas of 200 - 300 nm,[30] using the BCP pattern as a straightforward prototype for the investigation of the SIS process. In case that a specific registration of the pattern is required, the same procedure can be applied to pre-patterned substrates that can be used to externally drive the self-assembly process.[47,48]

Figure 3 shows representative FE-SEM plan view images of the BCP film with perpendicularly oriented lamellae before (a, f) and after exposure to $O_3$-based (b-e) and $H_2O$-based (g-j) SIS processes with different number of cycles ranging from one up to ten. In Figures 3(a, f) the BCP thin films were exposed to a short oxygen plasma process of 30 s to enhance the contrast in the FE-SEM images. The FE-SEM images of the infiltrated samples were acquired after an oxygen plasma step of 10 min that was performed to remove the polymeric template after infiltration.

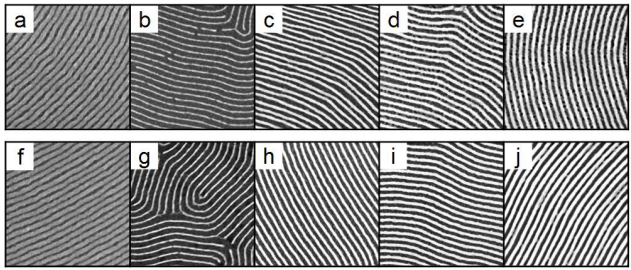

Figure 3. FE-SEM images of perpendicular PS-*b*-PMMA BCP lamellae before infiltration (a, f) and after infiltration (b-e and g-j). From b to e, BCP lamellae infiltrated with an $O_3$-based SIS process respectively for 1, 3, 5, and 10 SIS cycles. From g to j, BCP lamellae infiltrated with a $H_2O$-based SIS process for the same series of incremental SIS cycles.

From software analysis of the FE-SEM images, the characteristic dimensions of the lamellar features were determined. In Figure 4 the average width, height, and spacing obtained from this analysis are reported as a function of the number of SIS cycles. When only one SIS cycle is performed, very narrow alumina nanostructures are observed after polymer removal. The average width is slightly larger than 6 nm for both oxygen precursors. This value is well below the original size of the scaffolding PMMA domains which was measured to be 13 nm. As previously noticed, this provides additional flexibility for tuning of the final $Al_2O_3$ feature size at dimension smaller than the original BCP template and, in this case, well below 10 nm.[1] Starting from the third cycle, the lamellar width and height increase linearly with the SIS cycles, while the average spacing remains fairly constant and fixed at 26 nm since it is intrinsically determined by the choice of the BCP molecular weight that defines the lamellar pattern periodicity.[49] After ten consecutive SIS cycles, the average width reaches 13.5 nm, which is slightly larger than the original scaffold size, while the height stands at 16.5 nm. Setting aside the first cycle, in all cases the aspect ratio remains slightly above 1.2, which means that no substantial differences can be found between the growth along the lateral and vertical directions. Interestingly the growth trend closely reflects the one that was observed in the case of SIS in the 30 nm thick PMMA films that is characterized by a limited growth during the first cycle and a linear growth of the film thickness starting from three SIS cycles, with a slope of 0.40 ± 0.01 nm/cycle. In particular the thickness of the $Al_2O_3$ lamellae obtained after ten SIS cycles is equivalent to the thickness of the $Al_2O_3$ film obtained by SIS in the 30 nm thick flat PMMA film. This means that the final thickness of the $Al_2O_3$ structures is closely related to the initial thickness of the polymer template irrespective of its morphology.

The linear increase of the dimension as a function of the number of SIS cycles points out that no major polymer degradation occurs during SIS infiltration using the $O_3$ reactant. By comparing in Figure 4 the average dimensions extracted from similar samples infiltrated with the $H_2O$-based SIS process, it is interesting to note that these sample have almost the same lamellar width and average height within the experimental errors.

In summary, SIS using $O_3$ as oxygen precursor in BCP thin films with perpendicularly oriented lamellae produces, after removal of the polymeric template, a pattern of well aligned alumina stripes which closely mimics the original BCP template. The number of SIS cycles can be adjusted to effectively tune the height and width of the produced features, while the spacing between adjacent lamellae is fixed by the BCP template. In addition, this choice of perpendicular BCP morphology allows achieving different heights and widths and an aspect ratio above one. The average size of the features produced using the $O_3$-based SIS process are equivalent to the ones obtained using $H_2O$ as oxidant reagent.

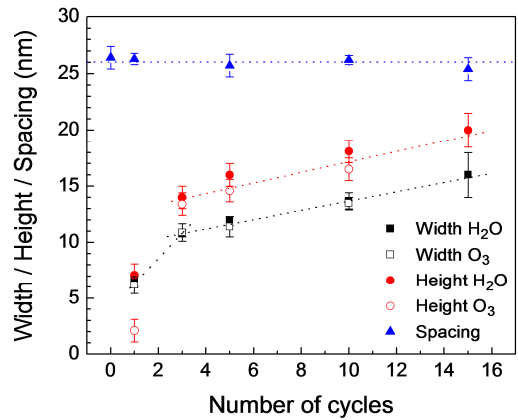

Figure 4. Average width, height, and spacing of the alumina lamellar features infiltrated with the $O_3$- and $H_2O$-based SIS processes as a function of the number of SIS cycles. Excluding the first cycle, the average width and height of the alumina lamellar features increases almost linearly with the number of SIS cycles for both processes, while the average spacing remains constant. Dotted lines are drawn as a guide to the eye.

### III.C. XPS analysis of the infiltrated PS-*b*-PMMA self-assembled lamellae

The alumina features obtained from BCP lamellae infiltrated with an increasing number of SIS cycles were inspected by XPS. In Figure 5, the survey spectra of the $Al_2O_3$ lamellae show the presence of Al and O core level signals. A clear indication of C contamination is present even after the removal of the polymeric template with oxygen plasma. For the sample obtained with one single SIS cycle XPS signals corresponding to silicon core levels are clearly visible due to the limited height of the alumina lamellae and to the consequent detection of photoelectrons coming from the Si underlying substrate.

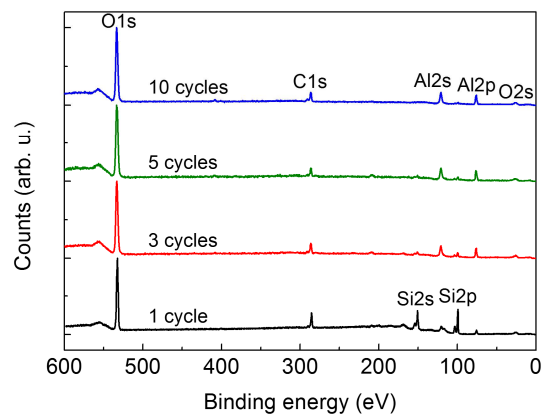

Figure 5. XPS survey spectra of the lamellar BCP infiltrated with the O$_3$-based process for different number of SIS cycles.

In Figure 6 the high-resolution spectra for the O 1s (left column) and Al 2p (central column) core level signals are shown. As previously discussed, the energy scales were aligned with respect to the Al 2p core level signal. Due to the limited thickness of the Al$_2$O$_3$ films after a few SIS cycles and to their not continuous nature, a relevant amount of signal in the O 1s spectra is coming from O atoms bind to Si in the native silicon oxide substrate. In particular, the O 1s signal coming from the substrate is a minority component in all spectra, except for one SIS cycle, in which ~60% of the whole O 1s signal comes from the silicon oxide. As reported in more details in the supplementary information S3, this contribution can be removed by proper and careful calibration of the spectra with XPS data obtained using a silicon substrate with native oxide as a reference. Figure 6 reports the O 1s high resolution spectra after proper removal of this components.

As in the case of the Al$_2$O$_3$ films obtained by SIS in flat PMMA films, the Al 2p core level signals (Figure 6, central column) can be fitted by a single narrow peak. Conversely the O 1s core level signals (Figure 6, left column) clearly display an asymmetric shape with a majority component at high BE. Interestingly, if we exclude the case of a single SIS cycle, all samples exhibit quite similar core level signals, indicating a stable oxide composition, irrespectively of the number of SIS cycles and the dimension of the alumina nanostructures.

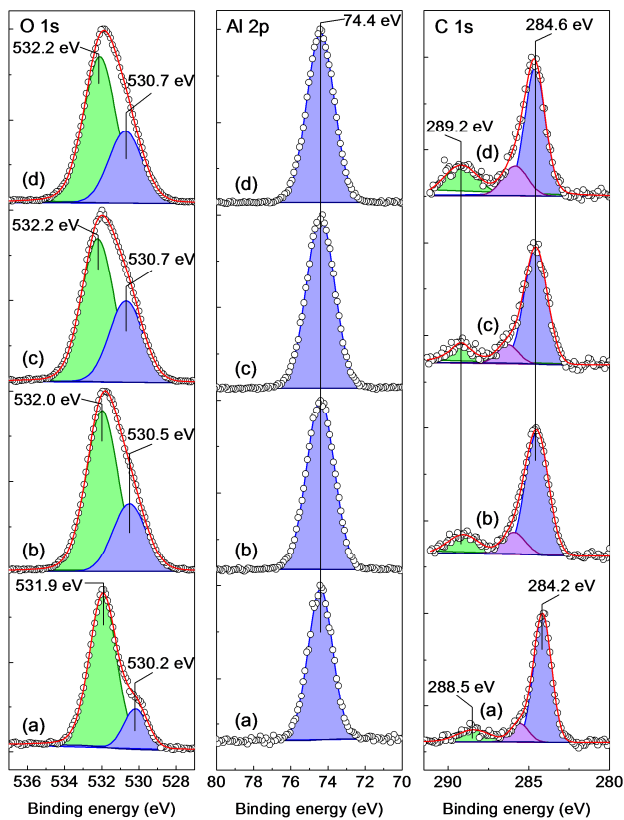

Figure 6. Normalized high-resolution XPS spectra of the O 1s, Al 2p, and C 1s portions of the spectrum acquired from BCP lamellae infiltrated with the O$_3$-based process for 1 cycle (a), 3 cycles (b), 5 cycles (c), and 10 cycles (d).

The same analysis was repeated on the alumina nanostructures created with the H$_2$O-based SIS process, obtaining similar results (Figure S4). The main difference concerns a little higher dispersion in the lower BE part of the O 1s spectrum. In this sense, O$_3$ seems to give more reproducible results among the various SIS cycles. The first cycle is in this case much more akin to the samples infiltrated with more SIS cycles. The chemical similarity obtained for the first cycle can be at least partially related to the higher height of the alumina lamellae after only one cycle using the H$_2$O-based process, as reported in Figure 4. The surface carbon content in samples obtained using the O$_3$-based SIS process is around 12-15%, slightly below the value found for the H$_2$O-based process and consistent with adventitious contamination.[7]

As previously noticed, an alumina composition closer to its stoichiometric value would require a deposition at higher temperature, similarly to what happens for ALD growth. However, a post deposition annealing can be used to remove the components related to contamination in the O 1s core level signals and obtain an O/Al ratio close to the stoichiometric alumina by moisture desorption and oxide densification. The Al$_2$O$_3$ lamellae were subjected to the same annealing procedure already performed on the flat alumina film (5 min at 400°C in N$_2$ ambient). The high resolution spectra of the O 1s, Al 2p, and C 1s core level signals obtained after annealing are reported in Figure 7. The O 1s peak shape is similar to the one reported in Figure 2, with the main component at low BE located at a distance of 456.4 eV with respect of the Al 2p peak. This component encompasses ~80% of the whole area and is compatible with stoichiometric alumina given the peak position and the O/Al ratio of ~1.5. Moreover, after annealing the surface carbon content dropped below 10% while the Al 2p signal did not change significantly.

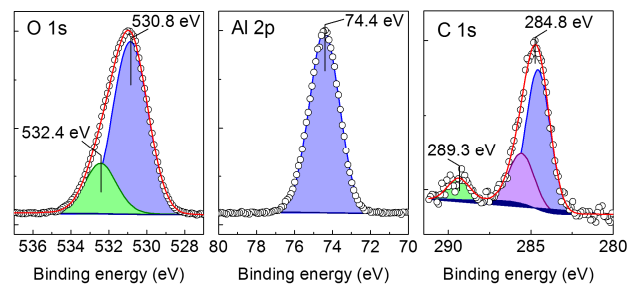

Figure 7. Normalized core level O 1s, Al 2p, and C 1s spectra of the infiltrated BCP lamellae (ten O$_3$-based SIS cycles) after annealing at 400°C for 5 min in N$_2$ ambient. The asymmetric O 1s peak was fitted with two Gaussian peaks.

After the thermal treatment, no significant difference was noticed in the spatial arrangement of the alumina lamellae. However, the average width contracted by 10% (from 13.5 ± 0.5 nm to 11.9 ± 0.5 nm). A similar shrinking was observed in flat alumina films deposited by ALD at low temperature.[40] It is worth to mention that the same annealing procedure was performed on the lamellae infiltrated with the H$_2$O-based process, obtaining very similar results.

## IV. Discussion

Data herein reported allows a direct comparison between the XPS high-resolution spectra of the SIS processes with the two different oxidants. A picture with the high-resolution spectra of the inorganic nanostructures obtained after ten SIS cycles is reported in Figure S5. From the general overview of the data we notice that both processes produce similar alumina in terms of relative XPS binding energy and peak shape. In both processes at least 3 cycles are necessary to stabilize the oxide composition. Moreover, a chemical evolution can be detected in the $Al_2O_3$ structures as function of the number of SIS cycles, with a progressive shift of the O 1s component related to alumina towards higher binding energy values (Figure 6). This slight variation in the chemical composition well correlates with the different trend in morphology evolution that is in Figure 4. A possible explanation of this evolution can be found in the different nature of reactive sites available for TMA during the first cycle with respect to the subsequent cycles. Actually at the beginning, only carbonyl groups are available in the PMMA volume, while starting from the second cycle Al-O sites are available inside the polymer scaffold, enhancing the TMA trapping efficiency.

An interesting comparison can be also drawn between the flat and the nanostructured alumina films produced using the $O_3$-based process. When ten consecutive SIS cycles are compared, the O 1s normalized signals are almost coincident and the Al 2p core level signals are compatible within the experimental error. Moreover, the flat PMMA film infiltrated with 10 SIS cycles produces, after polymer removal, a 15 nm thick $Al_2O_3$ film. The $Al_2O_3$ film thickness is equivalent to the average height of the lamellar features obtained after ten SIS cycles in the self-assembled BCP thin film. This observation corroborates the idea that the SIS process is almost independent on the dimensionality (flat vs. lamellae) of the polymer scaffold.

In summary, the SIS process based on alternating TMA and $O_3$ precursors in self-assembled perpendicular BCP lamellae leads to very good $Al_2O_3$ nanostructures with morphological and chemical characteristics, similar to those of $Al_2O_3$ films obtained in infiltrated flat PMMA films. Moreover, in all the aforementioned samples, the asymmetry of the O 1s peak is consistent with the one observed in alumina grown by ALD below 100°C. This asymmetry results in an excessive overall O/Al ratio, that is commonly attributed to hydration of the alumina surface by exposure to air.

It is important to remember that in this work, we focused on SIS in perpendicular lamellar morphology for both oxygen precursors. Several works in the literature report about $Al_2O_3$ nanostructures obtained by SIS in cylinder forming BCP thin films. Few works exist on the infiltration of BCP lamellae,[42,50] and no assessment is reported in the literature about the actual height and width of the resulting $Al_2O_3$ lamellae as a function of the number of SIS cycles. In this work we demonstrate that using lamellae forming BCP thin films, it is possible to obtain $Al_2O_3$ nanostructures exhibiting a similar range of accessible widths as compared to those obtained in case of BCP films with parallel cylinders. Moreover, the aspect ratio of the $Al_2O_3$ nanostructured obtained by SIS in lamellae forming BCP thin films is well above one, while in the case of BCP films with parallel cylinders the aspect ratio of the final $Al_2O_3$ nanostructures is usually limited to values lower than 0.5.[1,7] Preliminary results we obtained in BCP films with thickness higher than 30 nm, indicate that the height of the produced inorganic features is closely dependent on the initial thickness of the BCP film. Consequently, for the specific case of the perpendicular lamellar morphology, we expect to further increase the aspect ratio by increasing the thickness of the BCP film.[51,52]

From a general point of view infiltration of self-assembled BCP line patterns appears particularly interesting for pattern multiplication and densification in sub-10 nm scale lithography.[53] It is interesting to note that the alumina features obtained in this work have an average width ranging from $6.2 \pm 0.8$ nm to $13.5 \pm 0.5$ nm depending on the number of SIS cycles. These results are extremely promising even when compared with the typical dimensions obtained by infiltrating strongly segregating cylinder-forming BCP[46]. The lower dimension of the nanostructures obtained using these BCP implies, as a drawback, a much more complex processing of the polymeric as compared to PS-b-PMMA.[54] However considering the standard BCP phase diagram, in principle we could further reduce the degree of polymerization of PS-b-PMMA BCP preserving the ability to self-assemble in ordered nanostructures.[55] This would allow reducing the characteristic dimensions of the original self-assembled PS-b-PMMA template and generating, by properly tuning the SIS parameters, nanostructures with features well below 10 nm without the need for strongly segregating BCP.

## V. Conclusions

BCP self-assembled and PMMA thin films were infiltrated using a SIS process based on TMA and $O_3$ precursors. The resulting $Al_2O_3$ structures were compared with those obtained using a $H_2O$-based SIS process which stands as a model system for the SIS technique. In both cases XPS analysis revealed a chemical composition of the $Al_2O_3$ structures that is similar to the one observed in alumina films grown by ALD at low (< 100°C) temperature. No differences were noticed between the $Al_2O_3$ grown in the PMMA films and in the lamellae forming BCP thin films. The SIS processes using the two different oxygen precursors yielded alumina nanostructures with similar morphology, average dimension, and composition. Moreover, the adoption of the lamellae forming BCP appears promising for the development of nanostructured features with aspect ratio higher than one, irrespective of the oxygen precursor. In conclusion the result highlights the possibility to use $O_3$ as oxidant reactant for the SIS process, introducing an oxygen precursor characterized by high reactivity even at low temperature. This will allow expanding the possible range of metal precursors for the formation of oxide nanostructures that mimic the nanoscale pattern formed by self-assembled BCP templates.

## ASSOCIATED CONTENT

**Supporting Information**. AFM morphological analysis of the alumina lamellae; thickness of the infiltrated RCP as a function of the number of SIS cycles; detailed description of the procedure performed for the removal of the contribution from the substrate in the O 1s region of the XPS spectra; XPS core level spectra for different SIS cycles of the alumina lamellae produced using the $H_2O$-based SIS process; comparison of the XPS core level spectra of alumina lamellae produced using $O_3$ and $H_2O$ as oxygen precursor. This material is available free of charge via the Internet at http://pubs.acs.org.

## AUTHOR INFORMATION

**Corresponding Author**

*(J.F.) E-mail: jacopo.frascaroli@mdm.imm.cnr.it
*(M.P.) E-mail: michele.perego@cnr.it

**Author Contributions**

The manuscript was written through contributions of all authors.

**Notes**

The authors declare no competing financial interest. Patent protection related to this work is pending (International Patent Application No. PCT/IB2014/061324).

**ACKNOWLEDGMENT**

The authors would like to acknowledge Alessio Lamperti for his assistance with XPS measurements and the fruitful discussion. Federico Ferrarese Lupi (INRiM, Italy) is also gratefully acknowledged for the fruitful discussion. This research has been partially supported by the project "IONS4SET". This project has received funding from the European Union's Horizon 2020 research and innovation program under grant agreement No 688072.

For table of contents (TOC) only

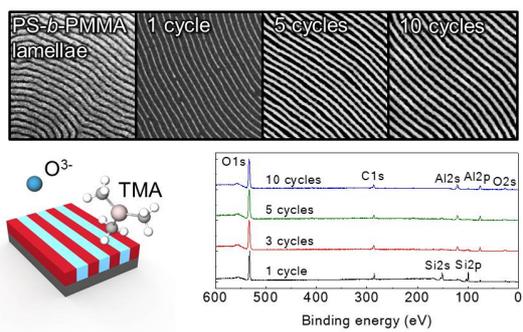